# Measuring granular flow through a funnel with a force sensor

Junghwan Lee, Kyeong Min Kim

*CheongShim International Academy, Gyeonggi-do, Republic of Korea*

**Abstract** – This paper presents an apparatus to measure the flow of granular matter through a funnel with a force sensor. In addition to being easily reproducible, our proposed method enhances measurement accuracy from the popular method that uses a stopwatch. We notice, however, that our proposed setup induces an unwelcome force that confounds force sensor readings. This issue is discussed, and its significance on measurement accuracy is evaluated through theoretical and experimental work.

## 1  Introduction

A well-known (and widely used) method of measuring the flow of granular matter through a funnel makes use of a stopwatch. Using the fact that granular flow through a funnel is constant [1], the rate of mass flow is determined by dividing the initial mass of the granular content inside the funnel by the time it takes to empty. While this method provides a great estimate of the flow rate, the time taken for the funnel to empty is recorded with a stopwatch, which relies on one's reaction time. This method therefore cannot be trusted for its accuracy. Hence, to provide an accurate alternative, we propose a method that makes use of a force sensor.

## 2  Apparatus and procedure

A schematic illustration and photo of our proposed apparatus is depicted in figure 1. To measure granular flow, the force sensor is first set to record the weight of the funnel and its granular content. Then, the funnel is opened, and its granular contents are left to flow out. While granular matter flows out the funnel, the force sensor will detect a linear decrease in weight. Then after the funnel has emptied, the recorded data is divided by gravitational acceleration (g) and a linear regression is posed. The slope of the fitted line is then the rate of mass flow.

## 3  An issue with the method

Unfortunately, however, the method that we propose induces a measurement error. As its granular content flows out of the funnel, the funnel's center of mass (CM) accelerates downward. As a result, an additional force, caused by CM acceleration, acts in addition to the weight of the funnel and confounds force sensor recordings into a mix of the two forces when weight was the only force we wanted to observe.

This phenomenon, expressed mathematically, is as follows:

$$\dot{m}_{Measured} = \dot{m}_{Weight} + \dot{m}_{CM} \quad (1)$$

where $\dot{m}_{Measured}$ is the mass flow measured by the force sensor, $\dot{m}_{Weight}$ is the detected mass flow due to the weight change, and $\dot{m}_{CM}$ is the mass flow detected due to CM acceleration.

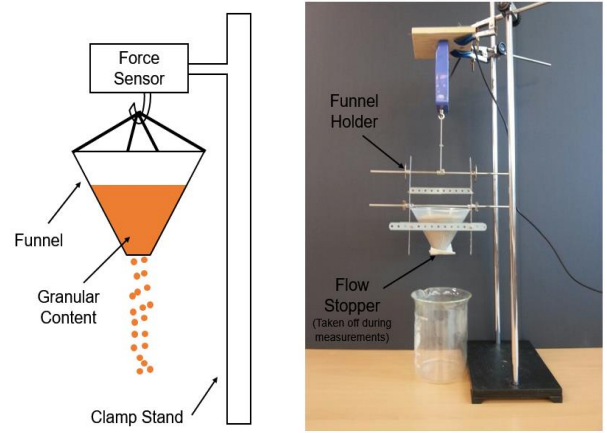

**Figure 1.** The apparatus.

Note that $\dot{m}_{Weight}$ is the real value of mass flow and $\dot{m}_{CM}$ is the error discussed above. In order for our method to give accurate measurements, $\dot{m}_{Measured}$ needs to equal $\dot{m}_{Weight}$, and this only happens when $\dot{m}_{CM}$ is negligible compared to $\dot{m}_{Weight}$.

To compare $\dot{m}_{CM}$ with respect to $\dot{m}_{Weight}$, we introduce $\zeta$, the ratio between $\dot{m}_{CM}$ and $\dot{m}_{Weight}$.

$$\zeta = \frac{\dot{m}_{CM}}{\dot{m}_{Weight}} \quad (2)$$

Finding the value of $\zeta$ will enable us to determine the significance of $\dot{m}_{CM}$ and tell us if the issue with our setup is negligible or not. To do so, we model for $\dot{m}_{CM}$ and $\dot{m}_{Weight}$.

To find $\dot{m}_{CM}$, we first start by modeling $F_{CM}$, which is the force caused by CM acceleration.

$$F_{CM} = (T - t)\dot{m}_{Weight} \cdot \frac{\partial^2}{\partial t^2}\left(\frac{1}{V}\int_0^H y \cdot A(y)\, dy\right) \quad (3)$$

In equation (3), the first term is the mass of the granular content, and the second term is its CM acceleration. Here, $T$ is the time it takes for the funnel to empty, $t$ the elapsed time since flow started, $V$ the instantaneous volume of the granular material inside the funnel, $H$ the height of the funnel, $y$ the vertical position, and $A(y)$ the cross-sectional area as a function of vertical position.

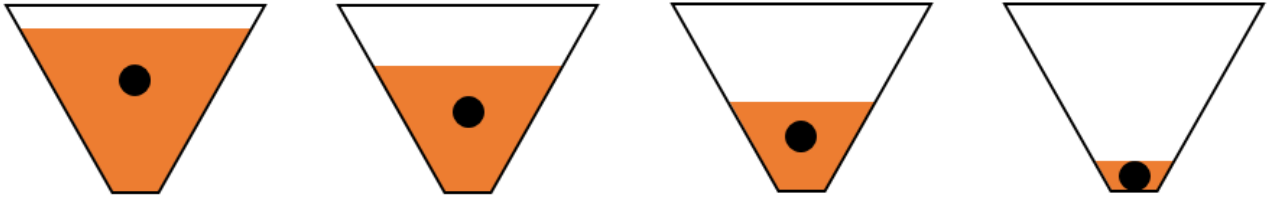

**Figure 2.** The center of mass (black) shifts downward as granular content (orange) flows out.

For a funnel,

$$A(y) = \pi \left(\frac{Rr_o - r_o}{H} \cdot y + r_o\right)^2 \quad (4)$$

and

$$V = \frac{\dot{m}(T-t)}{\rho} \quad (5)$$

where $r_o$ is the radius of the orifice, $\rho$ is the density of the granular content, and $R$ is the ratio between the longer radius and the shorter (orifice) radius. We plug in equations (4) and (5) into equation (3) and convert $F_{CM}$ into $\dot{m}_{CM}$ by taking a time derivative and dividing by $g$.

$$\dot{m}_{CM} = \frac{\pi \rho H^2 r_o^2 (1 + 2R + 3R^2)}{3g(T-t)^3} \quad (6)$$

And to model $\dot{m}_{Weight}$, we use a widely accepted law proposed by Beverloo et al. [1]

$$\dot{m}_{Weight} = C\rho(2r_o)^{2.5}\sqrt{g} \quad (7)$$

where $C$ is a dimensionless constant that depends on bulk density. However, as long as the orifice is considerably larger than a single granular particle, $C$ takes on a value near 0.5. [2]

Plugging in equations (6) and (7) into equation (2) gives

$$\zeta = \frac{\pi H^2 (1 + 2R + 3R^2)}{16.97\, C g^{1.5} r_o^{0.5} (T-t)^3} \quad (8)$$

Approximating $C$ as 0.5 and taking a time average of equation (8) to get rid of the time factor gives the average ratio $\bar{\zeta}$ as

$$\bar{\zeta} = \frac{4.19\, H^2 r_o^7 (1 + 2R + 3R^2)}{V_c^3} \quad (9)$$

where $V_c$ is the initial volume of the granular content.

If $\bar{\zeta}$ attains a value significantly low, e.g. under 0.01, then $\dot{m}_{Weight} \gg \dot{m}_{CM}$, and the issue with our method is negligible.

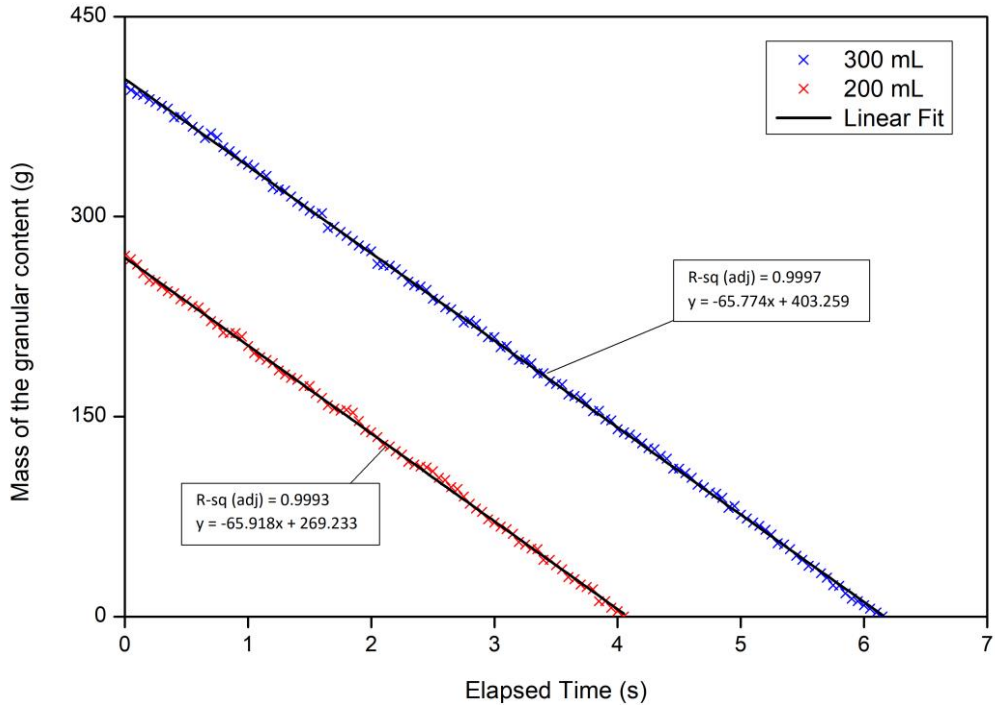

**Figure 3**: Mass of the granular content versus elapsed time, as measured by a force sensor.

## 4  Demonstration

We demonstrated our method on a PVC funnel ($R = 8.25$, $r_o = 0.8$ cm, and $H = 8.54$ cm) with 200mL and 300mL of dry sand (density = 1360 kg/m$^3$). Checking with equation (9) yielded $\bar{\zeta}$ as 0.0018 and 0.0005 for this funnel with 200mL and 300mL of dry sand, respectively. Because the value of $\bar{\zeta}$ is very small, we determined that, for our funnel, accurate measurements are produced.

The results of our demonstration are depicted in figure 3. We removed the force sensor readings at periods before and after the sand was flowing (when weight was constant), divided the recorded weight by $g$ and posed a linear regression on the scatterplot. As a result, both slopes, with a high r-squared value, gave the mass flow rate of this funnel as approximately 65.8 g/s. This result is in good agreement with the mass flow rate we found using a stopwatch, where an average of thirty trials gave 65.4 g/s.

## 5  Conclusion

Using a force sensor, we presented an apparatus to measure the flow of granular matter through a funnel. Concerns about inaccuracies due to CM acceleration were raised, and so we went on a journey to evaluate the issue. Applying our findings in equation (9) on our funnel determined the error to be negligible, and a demonstration further showed that our proposed method produces accurate measurements.

However, readers who wish to use this method are encouraged to plug-in parameters of their funnels into equation (9) before fully trusting the accuracy of the method.

## Acknowledgements

The authors are grateful to the Cheongshim International Academy Department of Science for providing us with lab space and equipment. Mr. Jaesung Yoon also gave comments that helped to improve our manuscript.